\documentclass[final,twocolumn,rmp,floats,amsfonts,amssymb,amsmath,superscriptaddress,showpacs,floatfix,a4paper]{revtex4}
\usepackage{graphicx}
\usepackage{hyperref}
\usepackage{subfigure}

\begin{document}
\title{Magnetic induction in a turbulent flow of liquid sodium: mean behaviour and slow fluctuations}
\author{F.~Ravelet}
\affiliation{Service de Physique de l'Etat Condens\'e, Direction des Sciences
de la Mati\`ere, CEA-Saclay, CNRS URA 2464, 91191 Gif-sur-Yvette cedex, France}
\affiliation{Present address: Laboratory for Aero and Hydrodynamics,
Leeghwaterstraat 21, 2628 CA Delft, The Netherlands}
\email{florent.ravelet@ensta.org}
\author{R.~Volk}
\author{A.~Chiffaudel}
\affiliation{Service de Physique de l'Etat Condens\'e, Direction des Sciences
de la Mati\`ere, CEA-Saclay, CNRS URA 2464, 91191 Gif-sur-Yvette cedex, France}
\author{F.~Daviaud}
\affiliation{Service de Physique de l'Etat Condens\'e, Direction des Sciences
de la Mati\`ere, CEA-Saclay, CNRS URA 2464, 91191 Gif-sur-Yvette cedex, France}
\author{B.~Dubrulle}
\affiliation{Service de Physique de l'Etat Condens\'e, Direction des Sciences
de la Mati\`ere, CEA-Saclay, CNRS URA 2464, 91191 Gif-sur-Yvette cedex, France}
\author{R.~Monchaux}
\affiliation{Service de Physique de l'Etat Condens\'e, Direction des Sciences
de la Mati\`ere, CEA-Saclay, CNRS URA 2464, 91191 Gif-sur-Yvette cedex, France}
\author{M.~Bourgoin}
\affiliation{Laboratoire de Physique de l'Ecole Normale
Sup\'erieure de Lyon, CNRS UMR 5672, 47 all\'ee d'Italie, 69364 Lyon 
Cedex 07, France}
\affiliation{Present address: Laboratoire des Ecoulements G\'eophysiques et Industriels, CNRS UMR 5519, BP53, 38041 Grenoble, France}
\author{P.~Odier}
\affiliation{Laboratoire de Physique de l'Ecole Normale
Sup\'erieure de Lyon, CNRS UMR 5672, 47 all\'ee d'Italie, 69364 Lyon 
Cedex 07, France}
\author{J.-F.~Pinton}
\affiliation{Laboratoire de Physique de l'Ecole Normale
Sup\'erieure de Lyon, CNRS UMR 5672, 47 all\'ee d'Italie, 69364 Lyon 
Cedex 07, France}
\author{M.~Berhanu}
\affiliation{Laboratoire de Physique Statistique de l'Ecole Normale
Sup\'erieure, CNRS UMR 8550, 24 Rue Lhomond, 75231 Paris Cedex 05, France}
\author{S.~Fauve}
\affiliation{Laboratoire de Physique Statistique de l'Ecole Normale
Sup\'erieure, CNRS UMR 8550, 24 Rue Lhomond, 75231 Paris Cedex 05, France}
\author{N.~Mordant}
\affiliation{Laboratoire de Physique Statistique de l'Ecole Normale
Sup\'erieure, CNRS UMR 8550, 24 Rue Lhomond, 75231 Paris Cedex 05, France}
\author{F.~P\'etr\'elis}
\affiliation{Laboratoire de Physique Statistique de l'Ecole Normale
Sup\'erieure, CNRS UMR 8550, 24 Rue Lhomond, 75231 Paris Cedex 05, France}

\begin{abstract}
We study the flow response to an externally imposed homogeneous magnetic field in a turbulent swirling flow of liquid sodium -- the VKS2 experiment in which magnetic Reynolds numbers $R_m$ up to 50 are reached. Induction effects are larger than in the former VKS1 experiment \cite[]{marie2002,bourgoin2002}. At $R_m$ larger than about 25, the local amplitude of induced field components supersedes that of the applied field, and exhibits non-Gaussian fluctuations. Slow dynamical instationarities and low-frequency bimodal dynamics are observed in the induction, presumably tracing back to large scale fluctuations in the hydrodynamic flow. 
\end{abstract}

\maketitle

\section{\label{sec:intro}Introduction}

The induction of a magnetic field by turbulent flows of electrically conducting liquids has received a great interest over the last decade, motivated by a better understanding of astrophysical and geophysical dynamos ---i.e. the generation of a self-sustained magnetic field by the flow of a conducting fluid \cite{larmor19}. For instance, Mari\'e {\em et al.} \cite{marie2002}, Bourgoin {\em et al.} \cite{bourgoin2002} and P\'etr\'elis {\em et al.} \cite{petrelis2003} study the so-called $\alpha$- and $\omega$-effects. Spence {\em et al.} \cite{Forest2006a} recently evidence induction effects that can only be attributed to the temporal fluctuations of the flow. In the same experiment, Nornberg {\em et al.} \cite{Forest2006b} evidence periods of intermittent growth of the induced magnetic field. Volk {\em et al.} \cite{volkpof2006} study the long-time fluctuations of induction profiles in liquid gallium. Stepanov {\em et al.} \cite{stepanov2006} study the turbulent $\alpha$-effect in a non-stationary flow. Finally, Sisan {\em et al.} \cite{sisan2003} study the influence of the Lorentz force on the turbulent dissipation and on the mean flow. A common feature of these studies is the use of an apparatus in which the instantaneous flow significantly differs from its time average. An interesting open issue is to quantify the role of fluctuations in the magnetic induction effects. Indications can be obtained by comparing induction measurements for different configurations in the same general experimental set-up.  

This is one of the goals of the VKS (von K\'arm\'an sodium) experiment which studies the magnetohydrodynamic behavior of the sodium flow generated inside a cylinder by counter rotation of various impellers (figure~\ref{fig:schema}). In a first series of experiments in 2000-2002 ---VKS1---, external steady magnetic fields are applied either along the cylinder axis or perpendicular to it \cite{marie2002,bourgoin2002,petrelis2003}. Modifications of this original set-up are motivated by the realization that electrical boundary conditions play an important role in the strength and geometry of magnetic induction \cite{peyresq2003,marieepjb2003,bourgoin2004,ravelet2005} and that fluctuations, as already noted, may have a leading effect. In a second evolution of the set-up ---VKS2---, the flow cell is enclosed within a layer of sodium at rest and thick copper casing resulting in different electrical boundary conditions. 

This article presents the flow response to an externally imposed homogeneous magnetic field in the VKS2 experiment.
Both VKS1 and VKS2 set-ups are described and compared in section~\ref{sec:setup}. We report in section~\ref{sec:response} induction characteristics in the presence of an externally applied field and compare them to that of VKS1. We identify and study a bimodal dynamics, for which the field abruptly changes between two states and stay there for variable durations, resulting in a very long timescales dynamics. In these regimes, the probability density functions of some of the components of the induced field, particularly the one aligned with the applied field, are non-Gaussian. Complementary observations are made in section~\ref{sec:annulus} when a thin annulus is placed in the mid shear layer. Finally, we discuss in section~\ref{sec:fin} some implications of our results regarding the dynamo capacity of VKS flows. 

\section{\label{sec:setup}The VKS experimental set-up and configurations} 
\begin{figure}[ht]
\begin{center}
\includegraphics[clip,width=80mm]{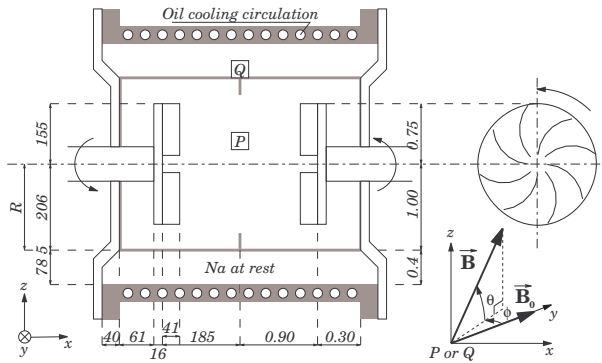}
\end{center} 
\caption{Sketch of the VKS2 experimental set-up. The inner and outer cylinders and the optional mid-plane annulus are made of copper (in gray). Other parts are stainless-steel. The dimension are given in millimeter (left) and normalized by the inner cylinder radius $R$ (right). The 3D Hall probe is located at point P, $0.25R$ from the axis. The angles characterizing the orientation of the induced field are displayed on the right. Magnetic measurements are made either at point P or at point Q when the annulus is present.}
\label{fig:schema} 
\end{figure}

Up to now, all VKS-experiments concern a swirling flow produced by two bladed facing impellers distant of $371$~mm in a $412$~mm diameter casing. The VKS1 flow region is bounded by a $10$~mm copper shell and a $20$~mm steel casing. The VKS2 evolution (figure~\ref{fig:schema}) basically consists of putting VKS1 inside a larger sodium volume and a $45$~mm copper casing, and designing new impellers. With respect to VKS1, the motor power has been increased from $150$~kW  to $300$~kW and the volume of the conducting domain is twice larger. The outer radial layer of sodium at rest surrounding the flow is of thickness $0.4R$. The counter-rotating impellers producing the flow are made of stainless steel, are of radius $0.75R$ and have 8 curved blades of $41$~mm height and $190$~mm curvature radius. They are labeled `TM73' in reference to the study of Ravelet {\em et al.} \cite{ravelet2005}. The velocity unit is based on the impellers rotation frequency $F$ and includes the stirring efficiency factor ${\cal V}=0.6$ of the TM73 impellers. This yields an integral Reynolds number $Re = {\cal V} 2\pi R^2 F /\nu$. The integral kinetic Reynolds number is of the order of $5\times10^6$. In addition, the shear-layer instability is a strong source of flow instationarity. 

Regarding the magnetic induction, the control parameter of the problem is the magnetic Reynolds number $R_m$ which compares the stretching of the magnetic field by velocity gradients to the magnetic diffusion. We define $R_m= {\cal V} \mu_0 \sigma 2\pi R^2 F$. The linear relation between the rotation frequency and the magnetic Reynolds number is $R_m\simeq1.88 F$ at $120^{\circ}$C, and $R_m\simeq1.73 F$ at $150^{\circ}$C. A temperature regulation has been installed, through oil circulating in the outer cylinder, in order to perform long time measurements in stationary regimes with temperature fixed in the range $110$ to $160^{\circ}$C (recall that the electrical conductivity of sodium varies significantly in the neighborhood of its melting temperature). The maximum magnetic Reynolds number is of the order of $50$, compared to about $35$ in VKS1. 

In order to study the MHD response of the flow, we apply a transverse field $\mathbf{B_0}=B_0 \mathbf{e_y}$ with a pair of coils. They are not in Helmholtz configuration and the inhomogeneity of $\mathbf{B_0}$, defined on the flow volume, can reach up to $20\%$ at the outer flow boundary. We measure the three components of the magnetic field $\mathbf{B}$ with a 3D-Hall probe set in the equatorial plane inside the flow, $50$~mm from the axis, {\em i.e.} at point $P$ in figure~\ref{fig:schema}. In this case, the closest VKS1 configuration \cite{marie2002} has impellers TM70 of radius $0.75 R$ and baffles on the cylinder wall. The applied field is less than 3~gauss, too weak a field to modify the flow: we have checked that the induced field varies linearly with the applied field.

\section{\label{sec:response}Response to an externally applied field}
Under the imposed field $B_0=2.7$~G, we study the dynamics of the three components of the dimensionless induced field $\mathbf{b}=(\mathbf{B}- \mathbf{B_0})/B_0$ and of its orientation at $P$. We define $\theta$ a latitude with respect to the $x-y$ plane, and $\phi$ the longitude in this plane, with respect to the applied field ($y$ axis) so that  $\mathbf{b}=||\mathbf{b}||\:[\cos\!\theta\,(-\sin\!\phi\,\mathbf{e_x} + \cos\!\phi \,\mathbf{e_y}) + \sin\!\theta\,\mathbf{e_z}]$. With these conventions, the applied field at $P$ is of unit norm with orientation $\theta=0$, $\phi=0$. 

We first present our observations about the mean magnetic induction properties of the flow and thereafter discuss the temporal dynamics of the signal.

\begin{figure}[htbp]
\begin{center}
\includegraphics[clip]{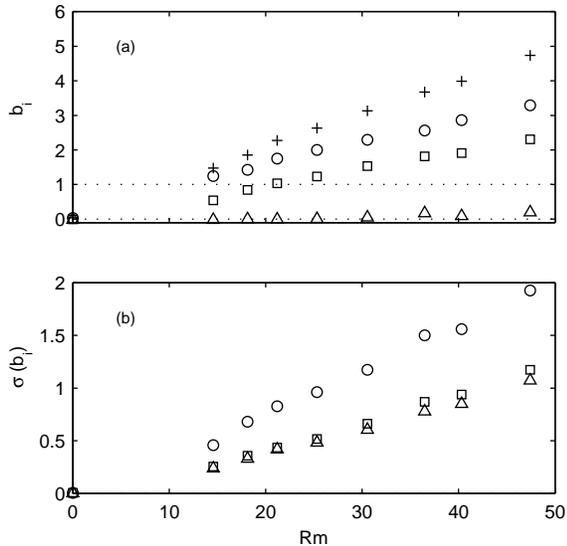}
\end{center} 
\caption{Mean value (a) and $rms$ amplitude (b) of the induced magnetic field {\em vs.} $R_m$, for the reduced magnetic field $\mathbf{b}=(\mathbf{B}- \mathbf{B_0})/B_0$ induced at point $P$. ($\circ$): $b_x$, ($\square$): $b_y$, ($\triangle$): $b_z$. and ($+$): $||\mathbf{b}||$.}
\label{fig:mean2} 
\end{figure}

\subsection{Mean induction}

We plot in figure~\ref{fig:mean2} the evolution of the mean (a) and standard deviation (b) of the components of $\mathbf{b}$. The mean vertical field $\langle b_z \rangle$ is zero, which can be explained by the symmetries of the time-averaged flow and of the applied field \cite{petrelis2003,bourgoin04mhd}. The induced field is dominated by the axial component $\langle b_x \rangle$ and once $R_m \geq 20$, the transverse component $\langle b_y \rangle$ becomes greater than the applied field. Due to limitations of the mechanical seals, we are limited to $F \geq 6$Hz ($R_m \gtrsim 11$). 

Results for the mean induction can be compared to the VKS1 experiment \cite{marie2002,bourgoin2002}. Recall that the main changes concern the electrical boundary condition due the added radial blanket of sodium, and the modification of the impellers. For $R_m > 12$, we observe that the behavior of $\langle \mathbf{b} \rangle(R_m)$ is linear, as in VKS1, but with a larger slope. We find $\langle b_x \rangle /B_0 \equiv R_m/R_m^{\star}$ with $1/R_m^{\star}=1/13$, a value $25 \%$ greater than in VKS1.   We also observe that the induced field no longer saturates for $R_m \gtrsim 25$: the induction is still growing at high $R_m$, leading to $\langle b_x \rangle \simeq 3.3 \ ; \ \langle b_y \rangle \simeq 2.3 $ at the highest $R_m \simeq 50$, while the corresponding VKS1 experiment saturates at $\langle b_x \rangle \simeq 1.5 \ ; \ \langle b_y \rangle \simeq 0.4$ for $R_m \gtrsim 25$. Actually, for $R_m \gtrsim 25$, the field component induced in the direction of the applied field is on average larger than the applied field although the system shows no sign of dynamo-self generation ---the induced field vanished as the applied field is turned off, save for induction due to the local Earth magnetic field. 

Finally, we observe that the standard deviations of the three components also increase linearly with $R_m$. Again there is no indication of a saturation for the larger $R_m$ values. The $rms$ amplitude are of the order of $50\%$ of the mean induced field. We note that for $R_m \gtrsim 45$, fluctuations in each component of the induced field exceed the magnitude of the applied field $\mathbf{B_0}$. In addition, the fluctuations of the axial component are twice larger than the fluctuations of the other components (figure~\ref{fig:mean2}b). In comparable VKS1 measurements the amplitude of fluctuations are the same for all components, with value $\sigma(b_i) \sim 0.5$ at $R_m \sim 25$. 

\subsection{Temporal and spectral dynamics}
\begin{figure*}[htbp!]
\begin{center}
\includegraphics[clip]{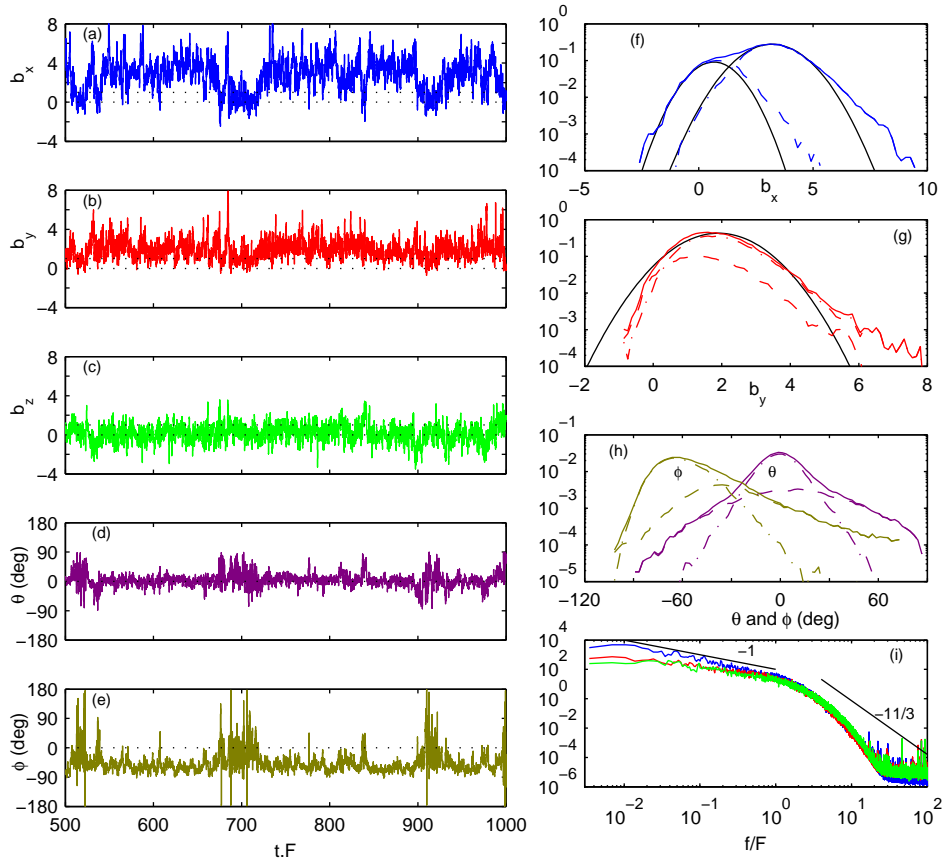}
\end{center} 
\caption{(a-e) Temporal signal of the components and of the orientations of $\mathbf{b}$ at $R_m=40$. 
(f,g) Probability density function (PDF) of $b_x$ and $b_y$ at $R_m=40$ (solid line). The dash-dotted line corresponds to the PDF in the $H-$states and the dashed line to the PDF in the $L-$states ---see section\,\ref{sec:bimodal} for details about the separation. Solid black lines are Gaussians. (h) Same plots for orientation angles $\theta$ and $\phi$. 
(i) Time spectrum of $b_x$ (blue), $b_y$ (red) and $b_z$ (green) at $R_m=40$.}
\label{fig:sig&pdf} 
\end{figure*}

In order to describe the time dynamics of the magnetic induction, we focus on measurements  at $R_m=40$ ($F=22$~Hz). We plot in figure~\ref{fig:sig&pdf} the time series, probability density function (PDF) and spectra of the components of the induced field as well as the $\theta$ and $\phi$ field orientation angles. As could be inferred from the magnitude of the standard deviations, one observes strong fluctuations in the induced field. 

Whereas in VKS1 all components of the magnetic field exhibit almost Gaussian PDF, we now observe non-Gaussian statistics for two components: $b_x$ and $b_y$ (figure~\ref{fig:sig&pdf}f-g). The component aligned with the applied field $b_y$ (figure~\ref{fig:sig&pdf}b and~\ref{fig:sig&pdf}g) exhibits an exponential tail of the form $\Pi(b_y) \propto \exp(-1.39 \times b_y)$. Its centered and reduced PDF does not depend on $R_m$. Upon closer inspection of figure~\ref{fig:sig&pdf}a, one also detects bimodality in the time evolution of the $b_x$ component. Relatively long periods of high induced field ({\em e.g.} around $t=600$ and $800 F^{-1}$) are followed by periods when $b_x$ is around zero ({\em e.g.} around $t=700$ and $920 F^{-1}$). This bimodality is analyzed in section\,\ref{sec:bimodal}.

We also show in figure~\ref{fig:sig&pdf}i the time spectrum of the three components of $\mathbf{b}$. At high frequency, the fall-off of the power spectra is steeper than the $-11/3$ power law decrease expected for the magnetic dissipative range in a Kolmogorov-like turbulence \cite{moffatt61}, {\em i.e.} for frequencies $f \gtrsim {R_m}^{3/4} \cdot F \simeq 20F$ at $R_m=40$. This steeper slope was also observed in previous VKS1 measurements, although the expected Kolmogorov behavior was observed in von K\'arm\'an gallium flows stirred by rugose disks \cite{odier98}. 

At low frequencies, roughly between $F/60 \lesssim f \lesssim F$,  the spectrum behaves approximately as a power law with an exponent $-1$ for $b_x$. For the two other components, the exponent is of the order of $-0.5$. This type of  spectral behavior is indicative of long-time evolutions in the magnetic induction. We believe that it traces back to slow changes in the underlying hydrodynamic flow, for instance the chaotic dynamics of the azimuthal shear layer observed in water prototype flows \cite{marie2004,theseravelet,ravelet2008}. It is also observed for induction measurements in von K\'arm\'an gallium flows \cite{volkpof2006} and is already present in VKS1 measurements. 

\subsection{\label{sec:bimodal}Bimodal analysis: High-State and Low-State}

We consider the bimodal dynamics illustrated above on the axial component of the magnetic field. These jumps between two states, which arise irregularly in time, appear even clearer when one plots the time evolution of the orientation of the induced field $\mathbf{b}$, as in figure~\ref{fig:sig&pdf}d-e. In the periods during which the $b_x$ component is large, the field orientation has smaller fluctuations whereas during periods for which $b_x$ is almost null, the orientation has strong fluctuations. Hence, the PDF of the axial induced component exhibit a clear bimodal behavior shown in figure~\ref{fig:sig&pdf}f and the PDFs of the field orientation are peaked but with large wings (figure~\ref{fig:sig&pdf}h): the PDF of $\theta$ is symmetric around the mean value $0$, while the PDF of $\phi$ has a peak around $-60^o$, close to the mean orientation, but is skewed towards positive angles.

\begin{figure}[htbp!]
\begin{center}
\includegraphics[clip]{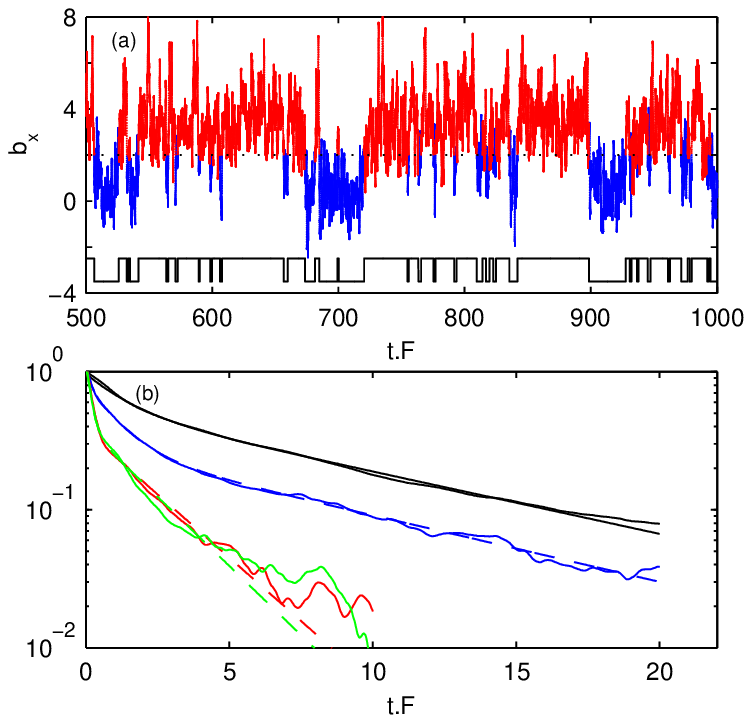}
\end{center} 
\caption{(a) Bimodal decomposition of $b_x$ component: $H-$state (red) and $L-$state (blue). Binary state signal (black).
(b) Autocorrelation functions in lin-log scale for the induced field components $b_x$ (blue), $b_y$ (red) and $b_z$ (green), with nonlinear fits by a sum of exponentials. Black solid line: autocorrelation function of the binary state signal. Measurements for $R_m = 40$; see text for details.}
\label{fig:bimodCorr} 
\end{figure}

In the following, we separate the time signals into two parts with a threshold criterion $b_{-}$ on $b_x$, illustrated in figure~\ref{fig:bimodCorr}a for $F=22{\rm Hz}$ ($R_m=40$) with $b_{-}=1.9$. We define the $L-$state (resp. $H-$state) as the subset when $b_x \le b_{-}$ (resp.$b_x>b_{-}$) during a time larger than $t_{min} \sim F^{-1}$. The remaining pieces are taken as $L-$state (resp. $H-$state) if the previous and the following signals are in the $L-$state (resp. $H-$state). This allows short magnetic field fluctuations below $b_{-}$ to belong to the $H-$state and vice versa. We also build a binary state signal with value $1$ in the $H-$states, and $0$ in the $L-$states (figure~\ref{fig:bimodCorr}a). 

We observe in figure~\ref{fig:sig&pdf}f that the PDF of $b_x$ in each state (dashed-dotted and dashed lines) are close to Gaussian fits of mean $0.66$ and standard deviation $1.21$ in the $L-$state with corresponding values  equal to $3.21$ and $1.59$ in the $H-$state (at $R_m=40$). Note that the shape of the PDF  of $b_y$ (figure~\ref{fig:sig&pdf}g) and $b_z$ (not shown) are not affected by the bimodal decomposition. The transverse component $b_y$ has its most probable value slightly shifted to a lower value in the $L-$state, with again exponential tails in both states ---a robust feature not affected by the bimodal decomposition. 
The PDF of the angles for the two separated states is plotted in figure~\ref{fig:sig&pdf}h. The field orientations are much more fluctuating in the $L-$state (standard deviation $\sigma (\phi)=34^{\circ}$) than in the $H-$state ($\sigma (\phi)=14^{\circ}$). The wide tail towards high values of $\phi$ belongs to the $L-$state. 

The bimodal analysis gives some insight into the slow evolution of the induction processes in the flow. It is best seen when one considers the autocorrelation functions of the induced components, i.e., another way to analyze the low frequency part of the power spectra. We plot the autocorrelation functions for $b_x$, $b_y$, $b_z$ in figure~\ref{fig:bimodCorr}b. They decrease exponentially at large time lags, and we perform a fit of these curves with a sum of exponentials. For $b_y$ ($b_z$ gives approximately the same results), the function is well fitted by two exponentials of decay times $2.4$ and $0.18 F^{-1}$. The $b_x$ component behaves differently and three exponentials are necessary to fit the curve. The corresponding time scales are $9.0$, $1.1$ and $0.11 F^{-1}$. The two last times are of the same order of magnitude as for $b_y$ and $b_z$: one is of the order of the disks period $F^{-1}$ and the shortest one is of the order of the period of a blade (since there are $8$ blades, the period is $0.125 F^{-1}$). However, $b_x$ shows a slow evolution with a time scale of the order of $9 F^{-1}$. This time arises from the bimodal transitions. Indeed, when we compute the autocorrelation for the binary state signal, we find that two exponentials are sufficient for the fit. Their characteristic times are $9.6$ and $1.3$ $F^{-1}$. We thus recover the two first times of $b_x$. The long time-scale, visible only for the binary state signal and for the complete $b_x$ signals, corresponds to $10$ times the impellers period and gives a bound corresponding to $f= 1/(2\pi \times 10 F^{-1}) \simeq F/60$ in the spectrum. We have checked for various $F$ that it seems to be a hydrodynamical time, of the order of magnitude of the coherence time scale of the largest coherent structures in the flow, created by the shear layer instability \cite{marie2004,ravelet2008}.

\begin{figure}[htp!]
\begin{center}
\includegraphics[clip]{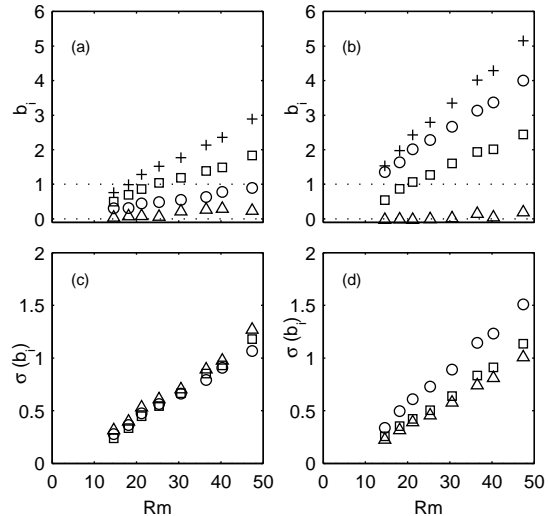}
\end{center} 
\caption{Evolution with $R_m$ for the $H$ and $L$-states. (a,b) mean value in the $L$-state and in the $H$ state; (c,d) same for the standard deviation. ($\circ$): $b_x$, ($\square$): $b_y$, ($\triangle$): $b_z$. and ($+$): $||\mathbf{b}||$.}
\label{fig:bimodRm} 
\end{figure}

Finally, we show in figure~\ref{fig:bimodRm} the evolutions with the magnetic Reynolds number $R_m$ of the mean and standard deviations for the induced magnetic field in the High- and Low-states. In each state, the mean values evolve linearly with $R_m$ but the components behave in a different way: in the $H-$state (figure~\ref{fig:bimodRm}b), they are close to the mean of the total field, with $\langle b_x \rangle > \langle b_y \rangle$, whereas in the $L-$state $\langle b_y \rangle > \langle b_x \rangle$ (figure~\ref{fig:bimodRm}a). The fluctuations of the induced field (figure~\ref{fig:bimodRm}c-d) are of the same order of magnitude in both states and are isotropic in the $L-$state.

\section{Induction with an additional annulus in the shear-layer.}
\label{sec:annulus}

\begin{figure*}[htp]
\begin{center}
\includegraphics[clip]{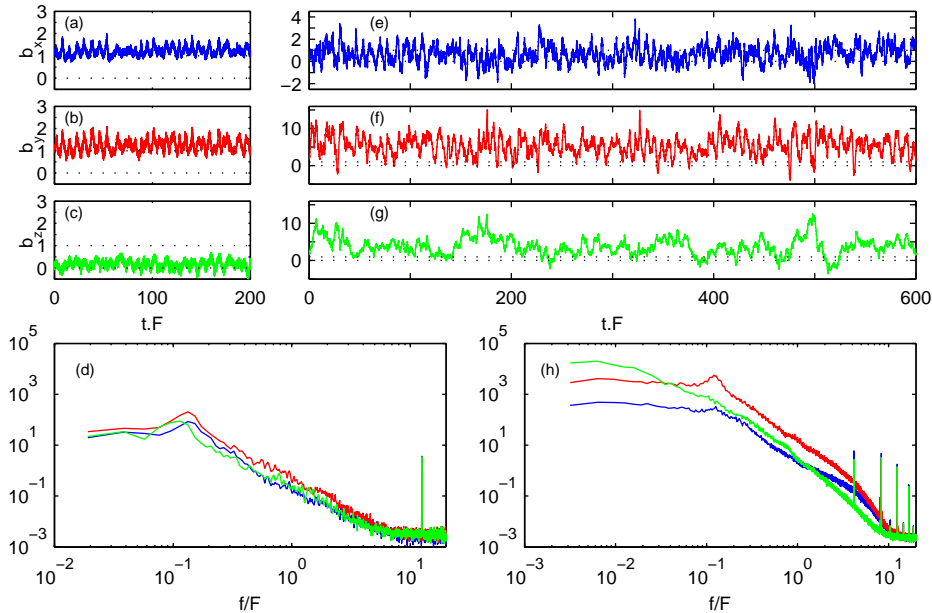}
\end{center} 
\caption{Temporal signals of the three components of the induced magnetic field and power spectra, with the annulus in the midplane. Measurements are done at point Q ---see figure~\ref{fig:schema}. $b_x$ (blue), $b_y$ (red) and $b_z$ (green). (a-d): $R_m=15$. (e-h): $R_m=42$. The dimensional fields are divided by the applied field measured at Q, {\em i.e.}, $B_0=1.8$ G.}
\label{fig:signalvks2f} 
\end{figure*}

Extensive testing and visualizations in water prototype experiments \cite{theseravelet} have shown that a way to stabilize the shear layer in the center of the flow is to introduce an annulus in the mid plane. Its leading effect is to reduce the low-frequency instationarities in the velocity field while the smaller scale fluctuations remain unchanged. As an illustration, the main structure of the free shear layer consists of three big fluctuating vortices. In the presence of an annulus, each vortex splits into a pair of smaller vortices that remain mostly attached to the leading edges of the annulus, one on each face. We made a series of sodium experiments for an annulus with  inner diameter $175$~mm inserted along the inner cylinder in the mid-plane between the disks. Magnetic measurements are then made at point $Q$, flush with the outer flow wall, just behind the annulus (figure~\ref{fig:schema}). The distance between points $P$ and $Q$ is of the order of several magnetic diffusion length, preventing complete quantitative comparisons. However, several interesting features emerge.

We show time series and time spectra of the induced magnetic field components for two $R_m$ in figure~\ref{fig:signalvks2f}. First, the time spectra of the $b_x$ and $b_y$ components level off for $f \lesssim 0.1 F$ indicating that several long-time dynamical features may have been suppressed, especially at low $R_m$. Instead, time-series and spectra do reveal a low-frequency oscillation between $F/10$ and $F/5$. This oscillation dominates the signal at low $R_m$ (figure~\ref{fig:signalvks2f}a-d). At higher $R_m$ (figure~\ref{fig:signalvks2f}e-h), this oscillation appears combined with lower frequency drifts ---especially on $b_z$. However, its dynamics is never comparable to the characteristic very-low-frequency binary jumps between $L-$state and $H-$state without annulus~: the fluctuations are indeed Gaussian ---in the sense that the PDF of all components $b_i(t)$ follow a Gaussian distribution.

\begin{figure}[htbp!]
\begin{center}
\includegraphics[clip]{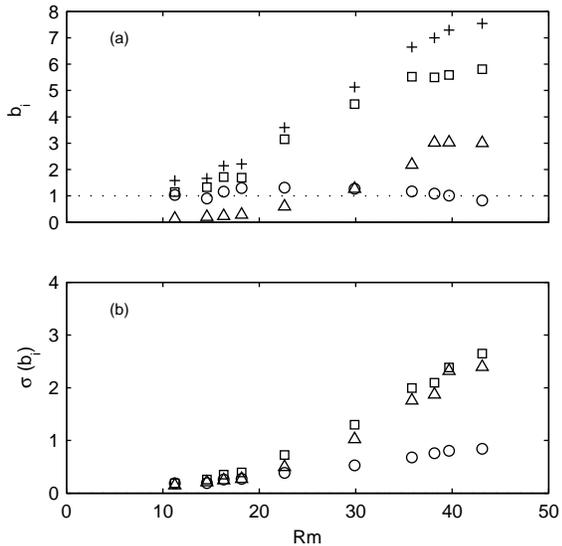}
\end{center} 
\caption{Mean value (a) and $rms$ (b) of the induced magnetic field {\em vs.} $R_m$, with the annulus in the midplane. ($\circ$): $b_x$, ($\square$): $b_y$, ($\triangle$): $b_z$ and ($+$): $||\mathbf{b}||$.}
\label{fig:moyvks2f} 
\end{figure}

We observe that the evolution of the mean induction also differs (figure~\ref{fig:moyvks2f}). The mean amplitude of $\mathbf{b}$ still depends linearly on $R_m$. It is dominated by the $b_y$ component ---parallel to $\mathbf{B_0}$. However, the evolution of $b_z$ and of the standard deviation of $b_y$ and $b_z$ shows a remarkable new feature compared to the previous induction measurements (including those performed in the VKS1 configuration): they are no longer linear function of $R_m$. Above $R_m \sim 30$ they start growing faster, together with the apparition of the very-low-frequency part of the $b_z$-spectrum, \emph{i.e.} below $F/10$.

\section{\label{sec:fin}Discussion and Concluding Remarks}

\emph{Comparison with VKS1 experiment.}
The measurements performed show that the addition of an surrounding layer of sodium and the modification of the driving impellers have lead to an increased induction efficiency as compared to former VKS1 runs. Regarding the boundary condition, this is in agreement with general considerations and numerical simulations: the effective magnetic Reynolds number is increased if currents can develop over a larger volume than the velocity domain \cite{bullard1977, avalos2003, marieepjb2003,bourgoin04mhd}. Regarding the flow generation, it shows the sensitivity of the induction to the precise geometry of the velocity gradients linked to the design of the driving impellers \cite{marieepjb2003,ravelet2005}. 

\emph{Effects of small appendices on temporal fluctuations in confined turbulent flows.}
Another finding is that the long-time dynamics observed here, as in many other turbulent flows in confined geometries, can be significantly altered by relatively small mechanical appendices, such as the thin annulus we placed in the equatorial plane (section~\ref{sec:annulus}). We indeed observe that, whereas the mean induction remains of the same order of magnitude, the slowest time-scale of the induced magnetic field is reduced by one order of magnitude with respect to the free case. This is in agreement with the hypothesis that such slow evolutions of the magnetic field are linked to large scale non-stationarities in the velocity field. In von K\'arm\'an flows, they originate mainly in the dynamics of the shear layer created by the counter-rotation of the driving disks, as evidenced in water prototype measurements \cite{theseravelet,ravelet2008} and induction measurements in liquid gallium ---for which the very low value of the magnetic Reynolds number result in the magnetic induction actually behaving as an image of velocity gradients \cite{volkpof2006}. 

\emph{Concluding remarks.}
Some of the observations described above may have implications regarding the self-generation of a magnetic field in a von K\'arm\'an flows. Dynamo predictions are often obtained from kinematic simulations in which the velocity field is time-independent, fixed to its time average value $\langle {\mathbf v} \rangle({\mathbf r})$.  In our case $\langle {\mathbf v} \rangle({\mathbf r})$ is measured from water model experiments. The threshold for which dynamo action is predicted varies significantly with prescribed boundary conditions, and, to a lesser extent, with the numerical procedure. Values range from $R_m^c=46$ for (finite axial boundary condition / $5$~mm copper shell separating the flow and the static conducting layer /copper container / no fluid behind the impellers) to $R_m^c > 100$ for the same previous conditions but with fluid in motion behind the impellers \cite{ravelet2005,stefani2006,nore2006}.  Numerical studies -- kinematic \cite{marieepjb2003} and dynamic \cite{thesejover} -- reveal that non linear trends should be detected when the magnetic Reynolds number reaches about $70\%$ of the dynamo threshold value. If this can directly be applied to the experiment, our results indicate that in the absence of the annulus, the critical magnetic Reynolds number, if it exists, should be larger than about $70$.  Note also that although we did observe that the induced magnetic field at the measurement location can greatly exceed the amplitude of the applied field (cf. figure~2), dynamo self-generation did not result. 

One related observation is that in order for dynamo generation to occur, several induction processes must cooperate. For instance, it has been shown that large scale phase fluctuations of the eddies of the G.~O.~Roberts flow \cite{petrelis2006}, or adding random large scale noise to the Taylor-Green flow \cite{laval06} could significantly increase the dynamo threshold (although natural Navier-Stokes fluctuations may actually decrease it compared to the kinematic value computed from the mean flow \cite{PontyNJP}. The introduction of an inner annulus, known to stabilize large scale fluctuations in the velocity field, did not lead to dynamo generation, but the measurements have shown a clear non-linear increase of the fluctuations of induction. 

It remains for future study to determine how crucial each of these factors is. Dynamo generation has been achieved in the VKS2 experiment \cite{monchaux2006} for a rather low magnetic Reynolds number, around $R_m=31$. The mechanical dynamo configuration corresponds to the second configuration (section~\ref{sec:annulus}) of the present article, {\em i.e.} with an annulus in the mid-plane, but the steel impellers used here are replaced by pure iron ones. Since these are of identical shape, the hydrodynamic flow is presumably the same (we checked that power consumption is identical in both flow below dynamo threshold) but the boundary conditions are quite different: the magnetic field does not penetrate the region behind the disks and magnetic field lines have to attach nearly perpendicular to the surface of the iron impellers. 

\begin{acknowledgments}

We acknowledge the assistance of  D.~Courtiade, C.~Gasquet, J.-B.~Luciani, P.~Metz, M.~Moulin, V.~Padilla, J.-F.~Point and A.~Skiara. We also would like to thank C.~Nore, R.~Laguerre, F.~Stefani and J.~L\'eorat for fruitful discussions. This work is supported by the french institutions: Direction des Sciences de la Mati\`ere and Direction de l'\'Energie Nucl\'eaire of CEA, Minist\`ere de la Recherche and Centre National de Recherche Scientifique (ANR 05-0268-03, GDR 2060). The experiments have been realized in CEA/Cada\-rache DEN/DTN.
\end{acknowledgments}

\bibliographystyle{unsrt}
\bibliography{BibVKS2a}

\begin{thebibliography}{10}

\bibitem{marie2002}
L.~Mari{\'e}, F.~P{\'e}tr{\'e}lis, M.~Bourgoin, J.~Burguete, A.~Chiffaudel,
  F.~Daviaud, S.~Fauve, P.~Odier, and J.-F. Pinton.
\newblock Open questions about homogeneous fluid dynamos: the {VKS} experiment.
\newblock {\em Magnetohydrodynamics}, 38:156--169, 2002.

\bibitem{bourgoin2002}
M.~Bourgoin, L.~Mari{\'e}, F.~P{\'e}tr{\'e}lis, C.~Gasquet, A.~Guigon, J.-B.
  Luciani, M.~Moulin, F.~Namer, J.~Burguete, A.~Chiffaudel, F.~Daviaud,
  S.~Fauve, P.~Odier, and J.-F. Pinton.
\newblock Mhd measurements in the von {K}\'arm\'an sodium experiment.
\newblock {\em Phys. Fluids}, 14:3046, 2002.

\bibitem{larmor19}
J.~Larmor.
\newblock How could a rotating body such as the sun become a magnet ?
\newblock {\em Rep. Brit. Assoc. Adv. Sci.}, page 159, 1919.

\bibitem{petrelis2003}
F.~P{\'e}tr{\'e}lis, M.~Bourgoin, L.~Mari{\'e}, J.~Burguete, A.~Chiffaudel,
  F.~Daviaud, S.~Fauve, P.~Odier, and J.-F. Pinton.
\newblock Nonlinear magnetic induction by helical motion in a liquid sodium
  turbulent flow.
\newblock {\em Phys. Rev. Lett.}, 90:174501, 2003.

\bibitem{Forest2006a}
E.~Spence, M.~Nornberg, C.~Jacobson, R.~Kendrick, and C.~Forest.
\newblock Observation of a turbulence-induced large scale magnetic field.
\newblock {\em Phys. Rev. Lett.}, 96:055002, 2006.

\bibitem{Forest2006b}
M.~Nornberg, E.~Spence, R.~Kendrick, C.~Jacobson, and C.~Forest.
\newblock Intermittent magnetic field excitation by a turbulent flow of liquid
  sodium.
\newblock {\em Phys. Rev. Lett.}, 97:044503, 2006.

\bibitem{volkpof2006}
R.~Volk, P.~Odier, and J.-F. Pinton.
\newblock Fluctuation of magnetic induction in von k\'arm\'an swirling flows.
\newblock {\em Phys. Fluids}, 18:085105, 2006.

\bibitem{stepanov2006}
R.~Stepanov, R.~Volk, S.~Denisov, P.~Frick, V.~Noskov, and J.-F. Pinton.
\newblock Induction, helicity, and alpha effect in a toroidal screw flow of
  liquid gallium.
\newblock {\em Phys. Rev. E}, 73:046310, 2006.

\bibitem{sisan2003}
D.~Sisan, W.~Shew, and D.~Lathrop.
\newblock Lorentz force effects in magneto-turbulence.
\newblock {\em Phys. Earth Planet Inter.}, 135:137--159, 2003.

\bibitem{peyresq2003}
S.~Fauve and F.~P\'etr\'elis.
\newblock The dynamo effect.
\newblock In J.-A. Sepulchre, editor, {\em Peyresq Lectures on Nonlinear
  Phenomena}, volume~2, pages 1--64. World Scientific, Singapore, 2003.

\bibitem{marieepjb2003}
L.~Mari{\'e}, J.~Burguete, F.~Daviaud, and J.~L{\'e}orat.
\newblock Numerical study of homogeneous dynamo based on experimental von
  {K}\'arm\'an type flows.
\newblock {\em Euro. Phys. J. B}, 33:469, 2003.

\bibitem{bourgoin2004}
M.~Bourgoin, P.~Odier, J.-F. Pinton, and Y.~Ricard.
\newblock An iterative study of time independent induction effects in
  magnetohydrodynamics.
\newblock {\em Phys. Fluids}, 16:2529, 2004.

\bibitem{ravelet2005}
F.~Ravelet, A.~Chiffaudel, F.~Daviaud, and J.~L\'eorat.
\newblock Toward an experimental von k\'arm\'an dynamo: Numerical studies for
  an optimized design.
\newblock {\em Phys. Fluids}, 17:117104, 2005.

\bibitem{bourgoin04mhd}
M.~Bourgoin, R.~Volk, P.~Frick, S.~Khripchenko, Ph. Odier, and J.-F. Pinton.
\newblock Induction mechanisms in von k\'arm\'an swirling flows of liquid
  gallium.
\newblock {\em Magnetohydrodynamics}, 40:13, 2004.

\bibitem{moffatt61}
H.K. Moffatt.
\newblock The amplification of a weak applied magnetic field by turbulence in
  fluids of moderate conductivity.
\newblock {\em J. Fluid Mech.}, 11:625, 1961.

\bibitem{odier98}
P.~Odier, J.-F. Pinton, and S.~Fauve.
\newblock Advection of a magnetic field by a turbulent swirling flow.
\newblock {\em Phys. Rev. E}, 58:7397--7401, 1998.

\bibitem{marie2004}
L.~Mari{\'e} and F.~Daviaud.
\newblock Experimental measurement of the scale-by-scale momentum transport
  budget in a turbulent shear flow.
\newblock {\em Phys. Fluids}, 16:457, 2004.

\bibitem{theseravelet}
Florent Ravelet.
\newblock {\em Bifurcations globales hydrodynamiques et
  magn\'etohydrodynamiques dans un \'ecoulement de von K\'arm\'an turbulent}.
\newblock PhD thesis, \'Ecole Polytechnique, 2005.

\bibitem{ravelet2008}
F.~Ravelet, A.~Chiffaudel, and F.~Daviaud.
\newblock Supercritical transition to turbulence in an inertially-driven von
  k\'arm\'an closed flow.
\newblock submitted to Journal of Fluid Mechanics, 2006.

\bibitem{bullard1977}
E.~C. Bullard and D.~Gubbins.
\newblock Generation of magnetic fields by fluid motions of global scale.
\newblock {\em Geophys. Astrophys. Fluid Dyn.}, 8:43, 1977.

\bibitem{avalos2003}
R.~Avalos-Zuniga, F.~Plunian, and A.~Gailitis.
\newblock Influence of electromagnetic boundary conditions onto the onset of
  dynamo action in laboratory experiments.
\newblock {\em Phys. Rev. E}, 68:066307, 2003.

\bibitem{stefani2006}
F.~Stefani, M.~Xu, G.~Gerbeth, F.~Ravelet, A.~Chiffaudel, F.~Daviaud, and
  J.~L\'eorat.
\newblock Ambivalent effects of added layers on steady kinematic dynamos in
  cylindrical geometry: application to the {VKS} experiment.
\newblock {\em Eur. J. Mech. B}, 25:894--908, 2006.

\bibitem{nore2006}
R.~Laguerre, C.~Nore, J.~L\'eorat, and J.-L. Guermond.
\newblock Effects of conductivity jumps in the envelope of a kinematic dynamo
  flow.
\newblock {\em C. R. M\'ecanique}, 334:593--598, 2006.

\bibitem{thesejover}
R.~Jover.
\newblock Induction in a turbulent {T}aylor--{G}reen flow.
\newblock Master's thesis, University Paris XI, 2006.

\bibitem{petrelis2006}
F.~P\'etr\'elis and S.~Fauve.
\newblock Inhibition of the dynamo effect by phase fluctuations.
\newblock {\em Eur. Phys. Lett.}, 76:602--608, 2006.

\bibitem{laval06}
J.-P. Laval, P.~Blaineau, N.~Leprovost, B.~Dubrulle, and F.~Daviaud.
\newblock Influence of turbulence on the dynamo threshold.
\newblock {\em Phys. Rev. Lett.}, 96:204503, 2006.

\bibitem{PontyNJP}
Y.~Ponty, P.~D. Mininni, J.-F. Pinton, H.~Politano, and A.~Pouquet.
\newblock Numerical study of dynamo action at low magnetic prandtl numbers:
  mean vs dynamical flow.
\newblock {\em New J. Phys.}, page submitted, 2007.

\bibitem{monchaux2006}
R.~Monchaux, M.~Berhanu, M.~Bourgoin, Ph. Odier, J.-F. Pinton, R.~Volk,
  S.~Fauve, N.~Mordant, F.~P\'etr\'elis, A.~Chiffaudel, F.~Daviaud,
  B.~Dubrulle, L.~Mari\'e, and F.~Ravelet.
\newblock Generation of a magnetic field by dynamo action in a turbulent flow
  of liquid sodium.
\newblock {\em Phys. Rev. Lett.}, 98:044502, 2007.

\end{thebibliography}

\end{document}